\documentclass[doublespacing]{elsart}

\usepackage{amssymb}
\usepackage{epsfig}

\begin{document}

\begin{frontmatter}



\title{Proposal for a spintronic femto-Tesla magnetic field 
sensor}


\author{S. Bandyopadhyay$^a$ and M. Cahay$^b$}
\address[label1]{Department of Electrical and Computer Engineering, 
Virginia Commonwealth 
University, Richmond, VA 23284, USA}
\address[label2]{Department of Electrical and Computer Engineering and Computer 
Science, University of Cincinnati, OH 45221, USA}

\begin{abstract}

We propose a spintronic magnetic field sensor, fashioned out of quantum wires,
which may be capable of detecting very weak magnetic fields with a sensitivity 
of $\sim$ 1 fT/$\sqrt{Hz}$ at a temperature of 4.2 K, and $\sim$ 80 
fT/$\sqrt{Hz}$ at 
room temperature. Such sensors have commercial applications in magnetometry, 
quantum computing, solid state nuclear magnetic resonance, 
magneto-encephalography, and military applications in weapon detection.

\end{abstract}

 \end{frontmatter}

\pagebreak
There is considerable interest in devising magnetic field sensors capable of 
detecting  weak dc and ac magnetic fields.  In this paper, we describe a 
novel concept for 
realizing such a device utilizing spin orbit coupling effects in a quantum wire.

Consider a semiconductor quantum wire with weak (or non-existent)
Dresselhaus  spin orbit interaction \cite{dresselhaus}, but a strong Rashba 
spin orbit interaction \cite{rashba} caused by an external transverse electric 
field. The Dresselhaus interaction accrues from bulk inversion asymmetry and is 
therefore virtually non-existent in centro-symmetric crystals, whereas the 
Rashba interaction arises from structural inversion asymmetry and hence can be 
made  large by applying a high symmetry breaking transverse electric field. We 
will 
assume that the wire is 
along the x-direction  and the external electric field inducing the 
Rashba effect is 
along the y-direction  (see Fig. 1).

This device is now brought into contact with the external 
magnetic field to be detected, and oriented such  that the field 
is directed along the wire axis (i.e. x-axis). Assuming that the field has a 
magnetic flux density $B$,
the 
effective mass Hamiltonian for the wire, in the Landau gauge 
${\bf A}$ = (0, $-Bz$, 0),  can be written as
\begin{eqnarray}
H & = & (p_x^2 + p_y^2 + p_z^2)/(2m^*) + (e B z p_y)/m^* + (e^2 B^2 z^2)/(2m^*) 
- (g/2) \mu_B B \sigma_x  \nonumber \\
& &  + V_y(y) + V_z(z) + \eta [ 
(p_x/\hbar) \sigma_z - (p_z/\hbar) \sigma_x ]
\end{eqnarray}
where $g$ is the Land\`e g-factor, $\mu_B$ is the Bohr magneton, $V_y(y)$ and 
$V_z(z)$ are the confining potentials along the y- and z-directions, $\sigma$-s 
are the Pauli spin matrices,  and $\eta$ is the strength of the 
Rashba spin-orbit 
interaction  in the wire. 

\begin{figure}[h]
\epsfxsize=4.3in
\epsfysize=4.3in
\centerline{\epsffile{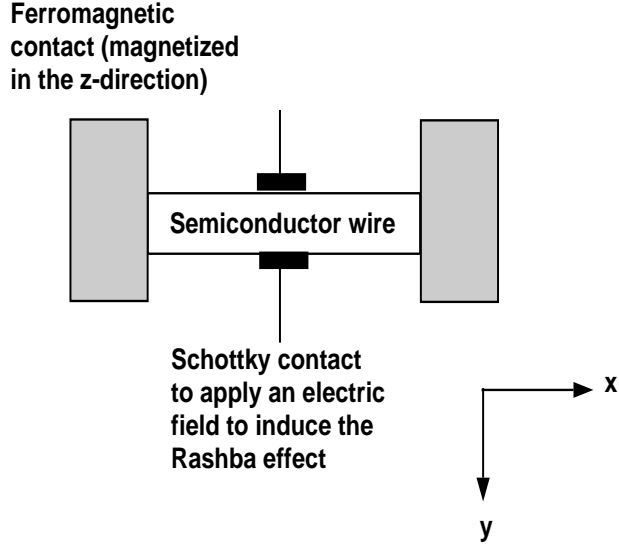}}
\caption[]{\small Physical structure of the magnetic field sensor. }
\end{figure}

We will assume that the wire is narrow enough and the temperature is low enough 
that 
only the lowest magneto-electric subband is occupied. In that case,
the Hamiltonian 
simplifies to  \cite{superlattice}
\begin{equation}
H = \hbar^2 k_x^2/(2 m^*) + E_0 -  
\beta 
\sigma_x + \eta k_x \sigma_z
\end{equation}
where $E_0$ is the energy of the lowest magneto-electric subband and $\beta$ = 
$g \mu_B B/2$.
   
Diagonalizing this Hamiltonian in a truncated Hilbert space spanning the two 
spin resolved states in the lowest subband yields the eigenenergies
\begin{equation}
E_{\pm} = {{\hbar^2 k_x^2}\over{2 m^*}}  + E_0 \pm 
\sqrt{ \left ( \eta    k_x \right )^2 + 
\beta^2} 
\label{eigenenergy}
\end{equation}
and the corresponding eigenstates
\begin{eqnarray}
{\Psi}_{+}(B, x) =
\left [ \begin{array}{c}
             cos(\theta_{k_x})\\
             sin(\theta_{k_x}) \\
             \end{array}   \right ]
             e^{i  k_x x}
             ~~~~~~~~~
{\Psi}_{-}(B, x) =
 \left [ \begin{array}{c}
              sin(\theta_{k_x})\\
              -cos(\theta_{k_x})\\
             \end{array}   \right ]
             e^{i  k_x x}
\label{eigenstate}
\end{eqnarray}
where $\theta_{k_x}$ = -$(1/2) arctan [\beta /\eta k_x]$. 

Note that if the magnetic flux density $B$ = 0, so that $\beta$ = 0, then
the energy dispersion relations given in Equation (\ref{eigenenergy}) are 
parabolic, but more importantly, the eigenspinors given in Equation 
(\ref{eigenstate}) are independent of $k_x$ because $\theta_{k_x}$ becomes 
independent of $k_x$. In fact, the eigenspinors become [1, 0] and [0, 1], which 
are +z-polarized and -z-polarized states. Therefore, with $B$ = 0,  each of 
the spin resolved subbands will have a definite spin quantization axis 
(+z-polarized and -z-polarized). 
Furthermore, these quantization axes are anti-parallel since the eigenspinors 
are orthogonal. 
As a result, there are no two states in the two spin-resolved subbands that can 
be coupled by any {\it non-magnetic} 
scatterer, be it an impurity or a phonon, or anything else. Hence, if a carrier 
is injected into the wire with its spin either +z-polarized or -z-polarized,
then the spin will not relax (or flip) no matter how frequently energy and 
momentum relaxing scattering events take place. The Elliott-Yafet mechanism of 
spin relaxation \cite{elliott} will also be completely suppressed since the 
eigenspinors are momentum independent. Furthermore, there is no D'yakonov-Perel' 
spin relaxation in a quantum wire if only a single subband is occupied 
\cite{ieee}. Therefore, {\it spin transport} will be 
ballistic when $B$ = 0, even if charge transport is not. In other words, the 
spin relaxation length would have been infinite were it not for such effects as 
hyperfine interaction with nuclear spins which can relax electron spin. Such 
relaxation can be made virtually non-existent by appropriate choice of 
(isotopically pure) 
materials. Even in materials that have isotopes with strong nuclear spin, the 
spin relaxation rate due to hyperfine interaction is very small.

Now consider the situation when the magnetic field is non-zero ($\beta \neq$ 0). 
Then the 
eigenspinors given by Equation (\ref{eigenstate}) are wavevector dependent. In 
this case, neither subband has a 
definite spin quantization axis since the spin state in either subband depends 
on the wavevector. Consequently, it is always possible to find two states in the 
two subbands with non-orthogonal spins. Any non-magnetic scatterer (impurity, 
phonon, etc.) can then couple these two states and cause a spin-relaxing 
scattering event. In this case,  no matter what spin eigenstate the carrier is 
injected in, spin transport is non-ballistic. That is to say, the spin 
relaxation length is much shorter compared to the case when there is no magnetic 
field. Since phonon scattering can flip spin in the presence of a magnetic 
field, and phonon scattering is quite strong in quantum wires (because of the 
van Hove singularity in the density of states) \cite{svizhenko1}, the spin 
relaxation length can be rather small in the presence of a magnetic field.
Therefore,  a magnetic field  drastically reduces the spin relaxation length. 
This is the basis of the magnetic field sensor.

The way the sensor works is as follows. Using a ferromagnetic spin 
injector contact magnetized in the +z-direction, we will inject carriers into 
the 
wire 
 with +z-polarized spins. The ferromagnetic contact results in a magnetic field 
directed mostly in the z-direction, but also with some small fringing component 
along the x-direction. The z-directed component of the field does not matter, 
since it does not extend along the wire, which is in the x-direction. However, 
any x-directed component will matter. Fortunately, the fringing field decays 
very quickly and if the wire is long enough, we can neglect the effect of the 
fringing field.   Therefore, in the 
absence of any {\it external} x-directed magnetic field, an injected electron 
will arrive with its spin polarization 
practically intact at the other end where another ferromagnetic contact 
(magnetized in the +z-direction) is 
placed. This second contact 
will then transmit the carrier completely and the spin polarized current will be 
high so that the device resistance will be {\it low}. However, if there is an 
external x-directed  
magnetic field, then the injected spin is no longer an eigenstate and therefore 
 can flip in the channel \cite{cahay_prb} and arrive at the second contact with 
arbitrary 
polarization. In fact, if the wire is long enough (much longer than the spin 
relaxation length), then the spin polarization of the current arriving at 
the second contact would have essentially decayed to zero, so that there 
is equal probability of an electron being transmitted or reflected. In this 
case, the device conductance can decrease by $\sim$ 50\% compared to the case 
when there is no magnetic field. This of course assumes that every carrier  was 
initially injected with its spin completely polarized in the +z-direction. In 
other words, the spin injection efficiency is 100\%. A more realistic scenario 
is to assume, say, a 32\% injection efficiency since it has already been 
demonstrated 
in an Fe/GaAs heterostructure \cite{hanbicki}. For 32\% injection efficiency, 
the conductance will decrease by 24\% in a magnetic field. We will err on the 
side of caution and assume conservatively that the device conductance will 
decrease by only 10\% in a 
magnetic field. A recent self-consistent drift diffusion simulation, carried out 
for a similar type 
of device,  has shown that the conductance decreases by 20 - 30\% because of 
spin flip scatterings 
\cite{saikin}.

At this point, we mention that the idea of changing device resistance by 
modulating spin flip scattering was the basis of a recently proposed ``spin 
field effect transistor'' \cite{schliemann}.  Unfortunately, the small 
conductance modulation achievable by this technique (maximum 50\%) is 
insufficient for a ``transistor'' which is an {\it active} device requiring a 
conductance modulation by three orders of magnitude to be useful. In contrast, 
the device proposed here is a {\it passive} sensor that does not have the 
stringent requirements of a transistor and, as we show below, a 10\% conductance 
modulation is adequate to provide excellent sensitivity.
 
Let us  now estimate how much magnetic field can decrease the device conductance 
by the assumed 10\%. Roughly speaking, this could correspond to the situation 
\begin{equation}
{{\beta}\over{\eta k_F}} \geq 0.1
\label{field}
\end{equation}
where $k_F$ is the Fermi wavevector in the channel. For a linear carrier 
concentration of 5$\times$10$^4$/cm, $k_F$ $\approx$ 8$\times$10$^6$/m. The 
measured value of $\eta$ in materials such as InAs is of the order of 10$^{-11}$ 
eV-m \cite{nitta1}. For most materials (with weaker Rashba effect), the value of 
$\eta$ may be two orders of magnitude smaller. Actually, this value can be tuned 
with an external electric field. Therefore, it is reasonable to estimate that 
$\eta k_F$ $\sim$ 1 $\mu$eV. The magnetic field that 
can cause the 10\% decrease in the conductance of the device is then found from 
Equation (\ref{field}) to be $\sim$ 10 Oe, if we assume $|g|$ $\approx$ 15.

We now carry out a standard sensitivity analysis following ref. \cite{tondra}.
The rms noise current for a single wire is taken to be the Johnson noise current 
given by
$I_n$ = $\sqrt{4kT G \Delta f}$ = $C \sqrt{G}$,
where $k$ is the Boltzmann constant, $T$ is the absolute temperature, $G$ is the 
device conductance, $\Delta f$ is the noise bandwidth, and $C$ = $\sqrt{4 kT 
\Delta f}$. In reality, the noise current can be suppressed in a quantum wire by 
two orders of magnitude  because of phonon confinement \cite{svizhenko1}. 
Therefore, $I_n$ = $C\sqrt{G}\zeta$, where $\zeta$ is the noise suppression 
factor which is about 0.01. There may be also other sources of noise, such as 
1/f noise, but this too is suppressed in quantum wires. 1/f noise can be reduced 
by operating the sensor under an ac bias with high frequency.

The change in current through a wire in a magnetic field $H$ is the signal 
current $I_s$ and is expressed as $SH$, where $S$ is the sensitivity. We will  
assume that the current drops linearly in a magnetic field, so that $S$ is 
independent of $H$. In reality, the current is more likely to drop superlinearly 
with magnetic field so that $S$ will be larger at smaller magnetic field. This 
is favorable for detecting small magnetic fields, but since we intend to remain 
conservative, we will assume that $S$ is independent of magnetic field. 
Therefore,
the signal to noise ratio for a single wire is
\begin{equation}
(S:N)_1 = {{SH}\over{I_n} } = {{S H}\over{C \zeta \sqrt{G}}}
\end{equation}

The noise current of $N$ wires in parallel is $C \sqrt {NG} \zeta$, whereas the 
signal current is $NSH$. Therefore, the signal-to-noise ratio of $N$ wires in 
parallel is $\sqrt{N}$ times that of a single wire.

For a carrier concentration of 5$\times$10$^4$/cm as we have assumed, and for a 
mobility of only 25 cm$^2$/V-sec (at 4.2 K),  a single wire of length 10 $\mu$m 
 has a conductance $G$ $\approx$ 2 $\times$ 
10$^{-10}$ Siemens. At a 100 mV bias, the current through a single 
wire is therefore 20 pA. Consequently, a 10\% change in conductance produces a 
change of 
current by 2 pA. Since the 10\% change in conductance is produced at 10 Oe, the 
{\it sensitivity} $S$ = 0.2 pA/Oe. At 1 fT (= 10$^{-11}$ Oe), the signal current 
$I_s$ for a single wire = $SH$ = 2 $\times$ 10$^{-24}$ A, whereas the noise 
current at a temperature of 4.2 K is 2 $\times$ 10$^{-18}$ A/$\sqrt{Hz}$. 
Therefore the signal to noise 
ratio for a single wire is 10$^{-6}$:1/$\sqrt{Hz}$ and that for a parallel array 
of 10$^{12}$ wires is 1:1/$\sqrt{Hz}$. Consequently, the optimum sensitivity is 
about 1 fT/$\sqrt{Hz}$ at 4.2 K. If we repeat the calculation for 300 
K (assuming that the mobility is reduced by a factor of 100 from its value at 
4.2 K), the optimum sensitivity is about 80 fT/$\sqrt{Hz}$.

We now calculate the power consumption of the sensor. The power dissipated in a 
single wire is 20 pA $\times$ 100 mV = 2 pW and for an array of 10$^{12}$ wires, 
the total power dissipation is 
2 W at 4.2 K. At room temperature, the conductance of each wire is reduced by 
approximately a factor of 100 because of mobility degradation, so that the power 
dissipation is reduced to 20 mW for the array. Therefore, the sensor we have 
designed is a low power sensor.

It should be obvious from the foregoing analysis that we can increase the 
sensitivity, without concomitantly increasing power dissipation, if we decrease 
the carrier concentration in the wire, but increase the mobility or decrease the 
wire length, to keep the conductance the same. This could allow sub-femto-Tesla 
field detection.

At a bias of 100 mV, the average electric field over a 10 $\mu$m long wire is 
100 V/cm. In the past, Monte Carlo simulation of carrier transport has shown 
that even at a much higher field than this, transport remains primarily single 
channeled in 
a quantum wire because of the extremely efficient acoustic phonon mediated 
energy relaxation \cite{svizhenko}. Therefore, the assumption of single 
channeled transport is mostly valid.

Finally, one needs to identify a realistic route to fabricating an array of 
10$^{12}$ quantum wires. We propose using a porous alumina template technique 
for this purpose. The required sequence of steps is shown in Fig. 2. A thin foil 
of aluminum is electropolished and then anodized in 15\% sulfuric acid at 25 V 
dc and at a temperature of 0$^{\circ}$C to produce a porous alumina film 
containing an ordered array of pores with 
diameter $\sim$ 25 nm \cite{nielsch}. A semiconductor is then selectively 
electrodeposited within the pores \cite{apl,menon} to create an array of 
vertically standing nanowires of 25 nm diameter. The density of wires is 
typically 10$^{11}$/cm$^2$, so that 10$^{12}$ wires require an area of 10 
cm$^2$. Next a ferromagnetic metal is electrodeposited within the pores 
\cite{menon1} on top of the semiconductor. The pores are slightly overfilled so 
that the ferromagnetic metal makes a two-dimensional layer on top. This is then 
covered with an organic layer to provide mechanical stability during the later 
steps.

The aluminum foil is then dissolved in HgCl$_2$, and the alumina barrier layer 
at 
the bottom of the pores is etched in phosphoric acid to open up the pores 
from the bottom. Another ferromagnet is then electrodeposited on the exposed 
tips of the semiconductor wires through a mask. The organic layer is then 
dissolved in acetone and the multilayered film is harvested and placed on top of 
a conducting substrate covered with silver paste.   Two 
 wires are attached to two remote diffused Schottky contacts in order to apply a 
transverse electric field that induces the Rashba effect. Finally wires are 
attached to top and bottom to provide ohmic electrical contacts. This completes 
the fabrication of the sensor. 

\begin{figure}[ht]
\epsfxsize=4.3in
\epsfysize=4.3in
\centerline{\epsffile{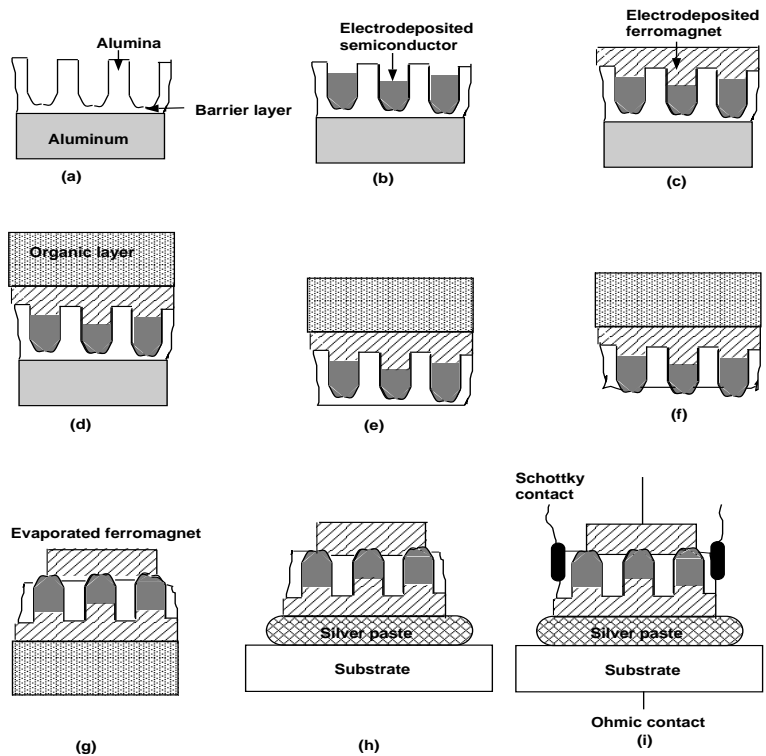}}
\caption[]{\small Fabrication steps. (a) Creation of a porous alumina film with 
25 nm diameter pores. This is produced by anodizing an aluminum foil at 25 V dc 
in sulfuric 
acid at 0$^{\circ}$C. (b) Electrodepositing a semiconductor selectively within 
the pores. (c) Electrodepositing a ferromagnetic layer on top. This could be 
either a metallic ferromagnet, or a semiconducting ferromagnet such as ZnMnSe. 
(d) Pasting a thick  organic layer on top for mechanical stability of the 
structure. (e) Dissolving out the Al foil in HgCl$_2$. (f) Etching the alumina 
in phosphoric acid at room temperature to expose the tips of the semiconductor 
wires. (g) Evaporating a ferromagnet on the tips of the semiconductor wires 
through a mask. (h) 
Dissolving the organic layer in ethanol, harvesting the film and placing it on 
top of a conducting substrate covered with silver paste. (i) Making Schottky 
contacts on the sides and ohmic contacts at top and bottom.}
\end{figure}

In conclusion, we have proposed a highly sensitive solid-state magnetic 
field detector based on spin orbit interaction. Such sensors can operate at 
room temperature since the energy separation between subbands can exceed the 
room temperature thermal energy in 25-nm diameter wires. The sensitivity of 
these 
sensors could be comparable to those of superconducting quantum interference 
devices \cite{gallup,barthelmess,pannetier} and therefore may be appropriate for 
magneto-encephalography where one directly senses the very small magnetic fields 
produced by neural activity in human brains.

The work of S. B. is supported by the Air Force Office of Scientific 
Research under grant FA9550-04-1-0261.

\pagebreak


\end{document}